\documentclass{appolb}
\usepackage{epsfig}
\begin{document}

\title{Off-diagonal generalized vector dominance in DIS and QCD.
\thanks{Presented by D. Schildknecht at the XXI School of Theoretical Physics, Ustro\'n, 
Poland, September 1999, to appear in Acta Physica Polonica.}
\thanks{Supported by the Bundesministerium f\"ur Bildung und
  Forschung, Bonn, Germany.}}

\author{G.~Cveti\v{c}$^{a,b}$\footnote{
cvetic@physik.uni-bielefeld.de},
D.~Schildknecht$^{a}$\footnote{
Dieter.Schildknecht@physik.uni-bielefeld.de} 
and A.~Shoshi$^{a}$\footnote{
shoshi@physik.uni-bielefeld.de}
\address{ $^a$Dept.~of Physics, University of Bielefeld ,
              D-33501 Bielefeld, Germany}
\address{ $^b$Dept.~of Physics, University of Dortmund,
              D-44221 Dortmund, Germany}
}






\maketitle

\begin{abstract}

We review the generalized vector dominance (GVD) approach to DIS at small values
of the scaling variable, $x$. In particular, we concentrate on a recent
formulation of GVD that explicitly incorporates the configuration of the
$\gamma^{\ast} \to q{\bar q}$ transition and a QCD--inspired ansatz for the
$(q{\bar q})p$ scattering amplitude. The destructive interference, originally
introduced in off-diagonal GVD is traced back to the generic strcuture of
two-gluon exchange. Asymptotically, the transverse photoabsorption cross section behaves as
$(\ln Q^2)/Q^2$, implying a logarithmic violation of scaling for $F_2$, while
the longitudinal-to-transverse ratio decreases as $1/\ln Q^2$. We also briefly
comment on vector-meson production.
\end{abstract}


\section{Introduction}

As a starting point for the present talk, I will briefly return to the experimental
results on photoproduction at high energies obtained in the late sixties and 
their interpretation in terms of vector meson ($\rho^0, \omega, \phi, J/\psi$) 
dominance (e.g. refs.~\cite{ref1, ref2}). I will subsequently introduce the concept of generalized 
vector dominance (GVD) \cite{ref3}, particularly relevant as soon as the photon four momentum enters the
spacelike region, $q^2=-Q^2<0$. I will briefly sketsch the basic points of the
most recent paper on off-diagonal GVD and its connection with QCD \cite{ref4}. Before entering the detailed discussions
in section 2, I will also present the motivation, essentially based on the results
obtained at
HERA \cite{ref5, ref6} on low--$x$ deep-inelastic scattering, for coming back to concepts that
existed already in the pre-QCD era. Essentially, it will be shown \cite{ref4} that
off-diagonal GVD, relevant in deep-inelastic scattering at low values of the
scaling variable $x \approx Q^2/W^2$, is very well compatible with and in fact
contained in QCD.

Experiments in the late sixties revealed that photoproduction \cite{ref1} from nucleons above
the resonance region has simple features characterized by
``hadronlike'' behaviour. Indeed, the energy dependence of the total cross section is much
like the energy dependence of typical hadron interactions, such as pion-nucleon
scattering. Likewise, as a function of the momentum transfer, the closely
related reaction of Compton scattering, $\gamma p \to \gamma p$, shows a forward
diffraction peak and has a predominantly imaginary forward scattering
amplitude. Similar features are observed in vector-meson production, $\gamma p
\to (\rho^0, \omega, \phi, J/\psi) p$. These empirical facts are quantitatively
summarized by vector-meson dominance \cite{ref2}. In intuitive terms, the photon virtually fluctuates
into the low-lying vector-meson states, which are subsequently scattered from
the nucleon. Upon applying the optical theorem, the total photoproduction cross
section is thus represented as a sum of vector-meson-forward-production
amplitudes \cite{ref7},
\begin{equation}
\sigma_{\gamma p}(W^2) = \sum_{V=\rho^0, \omega, \phi, J/\psi}
\sqrt{16\pi}\sqrt{\frac{\alpha\pi}{\gamma_V^2}} \left (
\frac{d\sigma^0}{dt}|_{\gamma p  \to Vp}(W^2) \right )^{\frac{1}{2}} \ ,
\label{eq1}
\end{equation}
and as a sum of total cross sections for the scattering of transversely
polarized vector mesons on protons \cite{ref8},
\begin{equation}
\sigma_{\gamma p}(W^2) = \sum_{V=\rho^0, \omega, \phi, J/\psi}
\frac{\alpha\pi}{\gamma^2_V} \sigma_{Vp}(W^2) \ .
\label{eq2}
\end{equation}
The factor $\alpha\pi/\gamma_V^2$ in (\ref{eq1}) and (\ref{eq2}) denotes the 
strength
of the coupling of the (virtual) photon to the vector meson $V$, as measured in
$e^+e^-$ annihilation by the integral over the vector-meson peak,
\begin{equation}
\frac{\alpha\pi}{\gamma^2_V} = \frac{1}{4\pi^2\alpha}\sum_F \int \sigma_{e^+e^-
  \to V \to F}(s) ds \ ,
\label{eq3}
\end{equation}
or by the partial width of the vector meson,
\begin{equation}
\Gamma_{V \to e^+e^-} = \frac{\alpha^2m_V}{12(\gamma^2_V/4\pi)} \ .
\label{eq4}
\end{equation}

The sum rule (\ref{eq1}) is an approximate one. The fractional contributions of
the $\rho^0, \omega$, and $\phi$ add up to approximately $72\%$ \cite{ref3, ref1}, and an
additional contribution of $1\%$ to $2\%$ is to be added for the $J/\psi$.

The lack of complete saturation of the sum rule (\ref{eq1}) in conjunction with
the coupling of the photon to a continuum of more massive states observed in
$e^+e^-$ annihilation provides the starting point of GVD. In obvious
generalisation of $\rho^0, \omega, \phi, J/\psi, \dots$ dominance, one assumes a double dispersion relation for the transverse photoabsorption cross 
section $\sigma_{\gamma^{\ast}_Tp}(W^2, Q^2)$ \cite{ref3},
\begin{equation}
\sigma_{\gamma^{\ast}_Tp}(W^2, Q^2) = \int dM^2 \int dM^{\prime 2}
\frac{\rho_T(W^2, M^2, M^{\prime 2})M^2M^{\prime 2}}{(Q^2+M^2)(Q^2+M^{\prime
    2})} 
\label{eq5}
\end{equation}
and a generalisation of (\ref{eq5}) to the longitudinal cross section,
$\sigma_{\gamma^{\ast}_Lp}(W^2, Q^2)$. The spectral weight function,
$\rho_T(W^2, M^2, M^{\prime 2})$, contains the coupling  stren-gths of the
(timelike virtual) photon to the hadronic vector states of masses $M$ and
$M^{\prime}$, as well as the (imaginary parts of the) forward-scattering
amplitudes of these states on the nucleon. The respresentation (\ref{eq5}) was
expected \cite{ref3} to be valid in particular in the small-$x$ diffraction region of deep
inelastic scattering (DIS), i.e. at small $x$ and all $Q^2$.\footnote{Compare
  ref.~\cite{ref8a} for related lifetime arguments.} 

\begin{figure}[htb] 
\unitlength 1mm
\begin{center}
\begin{picture}(126,145)
\put(2,2){
\epsfxsize=11cm
\epsfysize=13cm
\epsfbox{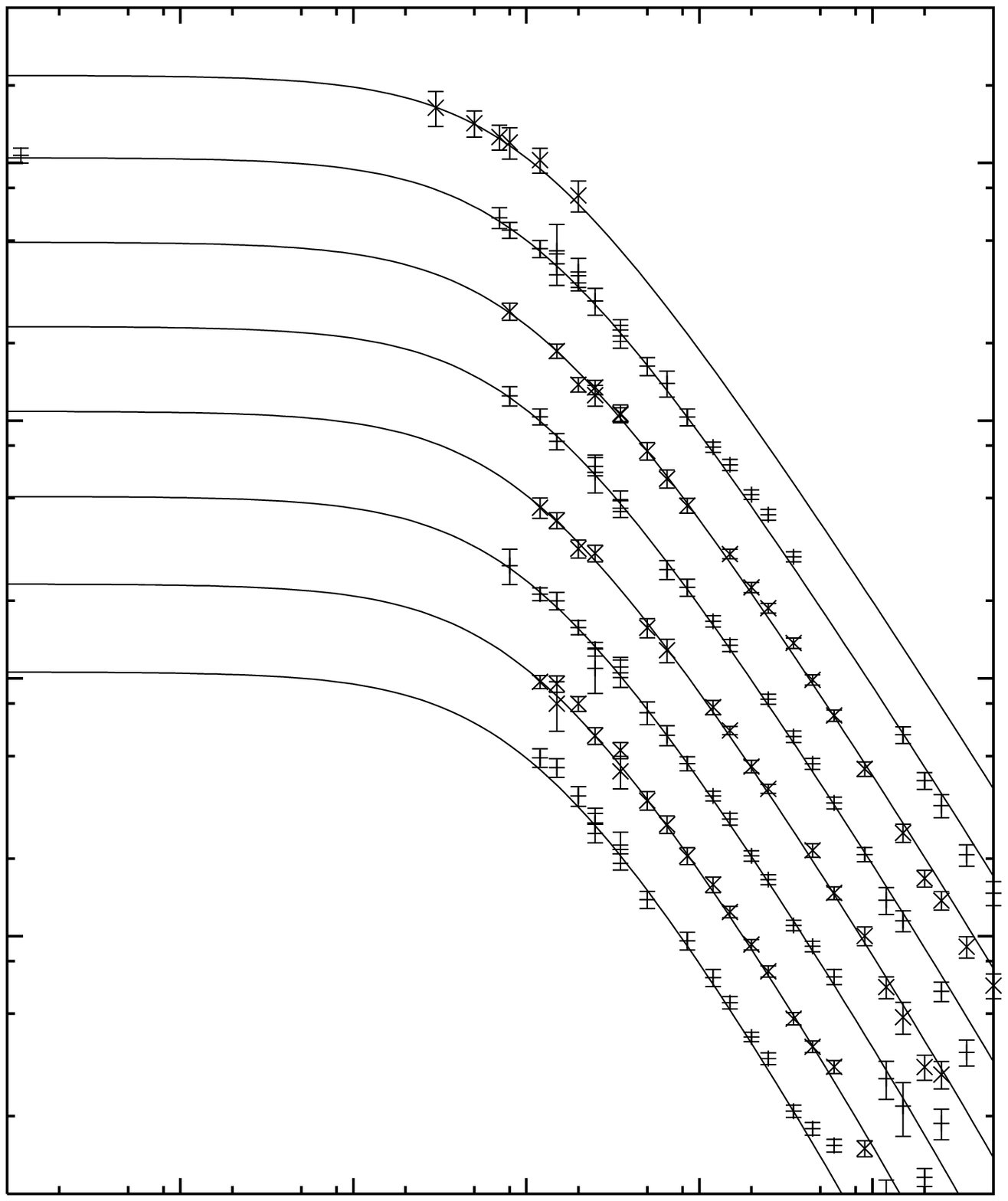}}
\put(2,2){
\epsfxsize=11cm
\epsfysize=13cm
\epsfbox{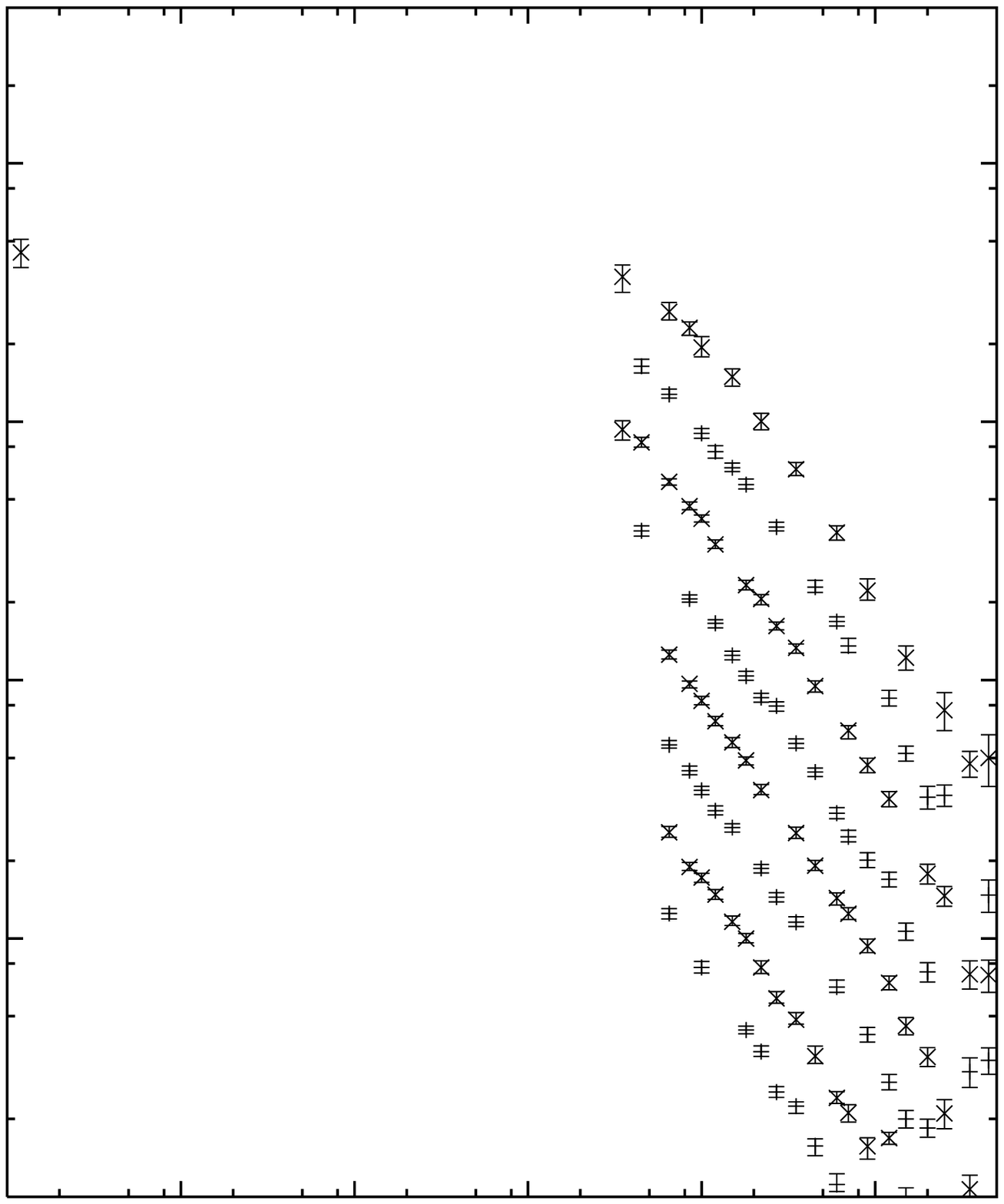}}
\put(100,0){\makebox(0,0){\large $Q^2$[GeV$^2$]}}
\put(99,6){\makebox(0,0){$10^2$}}
\put(81.6,6){\makebox(0,0){$10$}}
\put(64.2,6){\makebox(0,0){$1$}}
\put(47.8,6){\makebox(0,0){$10^{-1}$}}
\put(30.4,6){\makebox(0,0){$10^{-2}$}}
\put(13,6){\makebox(0,0){$10^{-3}$}}
\put(10.5,115){\makebox(0,0)[r]{$10^4$}}
\put(10.5,89){\makebox(0,0)[r]{$10^3$}}
\put(10.5,63){\makebox(0,0)[r]{$10^2$}}
\put(10.5,37){\makebox(0,0)[r]{$10$}}
\put(12,11){\makebox(0,0)[r]{$1$}}
\put(16,126){\makebox(0,0)[l]
{\large $\sigma_{\gamma^*p}[\mu{\rm b}]$}}
\put(16,120){\makebox(0,0)[l]{\small $\times 128,~W=245$ GeV}}
\put(16,111.4){\makebox(0,0)[l]{$\times 64,~W=210$ GeV}}
\put(16,102.85){\makebox(0,0)[l]{$\times 32,~W=170$ GeV}}
\put(16,94.3){\makebox(0,0)[l]{$\times 16,~W=140$ GeV}}
\put(16,85.7){\makebox(0,0)[l]{$\times 8,~W=115$ GeV}}
\put(16,77.1){\makebox(0,0)[l]{$\times 4,~W=95$ GeV}}
\put(16,68.6){\makebox(0,0)[l]{$\times 2,~W=75$ GeV}}
\put(16,60){\makebox(0,0)[l]{\small $W=60$ GeV}}
\end{picture}
\end{center}
\caption{
  Generalized Vector Dominance prediction for $\sigma_{\gamma^*p}$ \cite{ref11}
  compared with the experimental data from the H1 and ZEUS collaborations 
at HERA \cite{ref5, ref6}.}
\label{fig8}
\end{figure}

The, for simplicity, frequently employed diagonal approximation
\begin{equation}
\rho_T \sim \delta(M^2-M^{\prime 2})
\label{eq6}
\end{equation}
together with the photon coupling, i.e. $\sigma_{e^+e^- \to hadrons}(M^2) \sim
1/M^2$, requires that vector-state-nucleon scattering, $\sigma_{Vp}$, is to decrease as
$\sigma_{Vp} \sim 1/M^2$. Otherwise, the representation (\ref{eq5}) becomes
divergent, and scaling of the nucleon structure function becomes violated by a
power of $Q^2$. One arrives at what sometimes \cite{ref9} has been called the ``Gribov
paradox'' \cite{ref10}. Due to the existence of diffraction dissociation in hadron-induced
reactions, corresponding to off-diagonal transitions, the validity of the 
diagonal approximation was doubtful right from the
outset. The $1/M^2$ law for $\sigma_{Vp}$ has nevertheless been frequently
applied as an effective one. In the HERA energy range, it led to acceptable 
fits to the experimental data \cite{ref11}. Compare fig.~\ref{fig8}.

Taking into account the known empirical behaviour of diffraction dissociation
(i.e. $M \neq M^{\prime}$) in hadron physics, it was noted \cite{ref12} a long time ago that
indeed the justification for the diagonal approximation (\ref{eq6}) stands on
extremely weak grounds. By invoking destructive interference between diagonal
and off-diagonal transitions, motivated by quark-model arguments for
photon-vector-meson couplings, it was shown \cite{ref12}\footnote{Compare also \cite{ref9a}
  for a comparison of off-diagonal GVD with the experimental data then available.} that indeed scaling in $e^+e^-$
annihilation ($\sigma_{e^+e^-} \sim 1/M^2$) and a constant total vector-state-nucleon cross section ($\sigma_{VN} \sim$ const, independent of $M$) are
completely compatible with scaling ($\sigma_{\gamma_T^{\ast}p} \sim 1/Q^2$) in
DIS.

I finally turn to the most recent work \cite{ref4} by Cvetic, Shoshi and myself. In contrast to
the pre-QCD formulation of GVD, we explicitly take into account the
configuration of the 
$\gamma^{\ast} \to q{\bar q}$ transitions (seen in quark-jets in $e^+e^-$
annihilation) and a QCD-inspired structure for the $(q{\bar q})p$-forward-scattering amplitude. We will see that the destructive interference
introduced in off-diagonal GVD \cite{ref12} is recovered in this formulation and traced back
to the generic structure of two-gluon exchange. While the general structure of
the expression for the transverse and longitudinal cross sections is the same as
in the original formulation of off-diagonal GVD, there is nevertheless an
imptortant modification in terms of a $\log Q^2$ factor that originates
precisely from the explicit incorporation of the $q{\bar q}$ configuration in
the $\gamma^{\ast} \to q{\bar q}$ coupling. 
\begin{figure}[htb]
\setlength{\unitlength}{1.cm}
\begin{center}
\epsfig{file=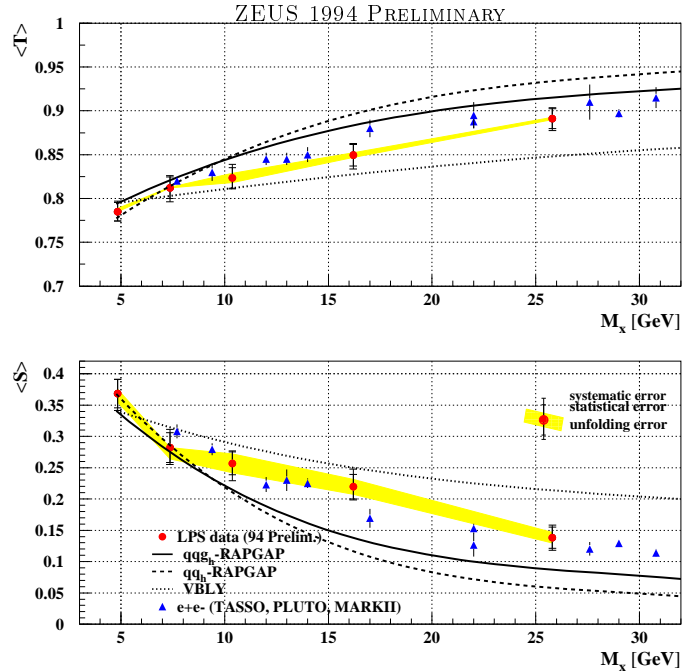,width=10.cm}
\end{center}
\caption{\small Thrust $<T>$ and sphericity $<S>$ in diffractive production and
$e^+e^-$ annihilation.}
\label{fig1}
\end{figure}
In what follows, as in the original
paper \cite{ref4}, theoretical issues will be stressed, while the comparison with the data
will at most be qualitative. A quantitative comparison with the data, which 
needs a more careful treatment of the energy dependence, will
hopefully be provided in the near future. 

Before coming to the main subject, a
brief remark on the motivation for taking up the subject of off-diagonal GVD
again may be appropriate. The motivation is essentially provided by the fact that HERA
is able to explore low--$x$ DIS in detail. Moreover, qualitatively, the
experimental HERA results are as expected from GVD and strongly support the 
GVD ansatz,

\begin{enumerate}
\item[(i)]
the existence of diffractive production of high-mass states, the
``large-rapidity-gap events'' first announced at the '93 Marseille 
International Conference \cite{ref13},
\item[(ii)]
the similarity in shape to $e^+e^-$ final states, compare fig.~\ref{fig1} from \cite{ref13a},
\item[(iii)]
the persistance of shadowing on complex nuclei in DIS, observed since about the
year 1988 \cite{ref14}, after many years of confussion.
\end{enumerate}

\section{Off-diagonal GVD from QCD}

\subsection{The classical GVD approach in momentum space}
\label{subs1}

\begin{figure}[htb]
\setlength{\unitlength}{1.cm}
\begin{center}
\epsfig{file=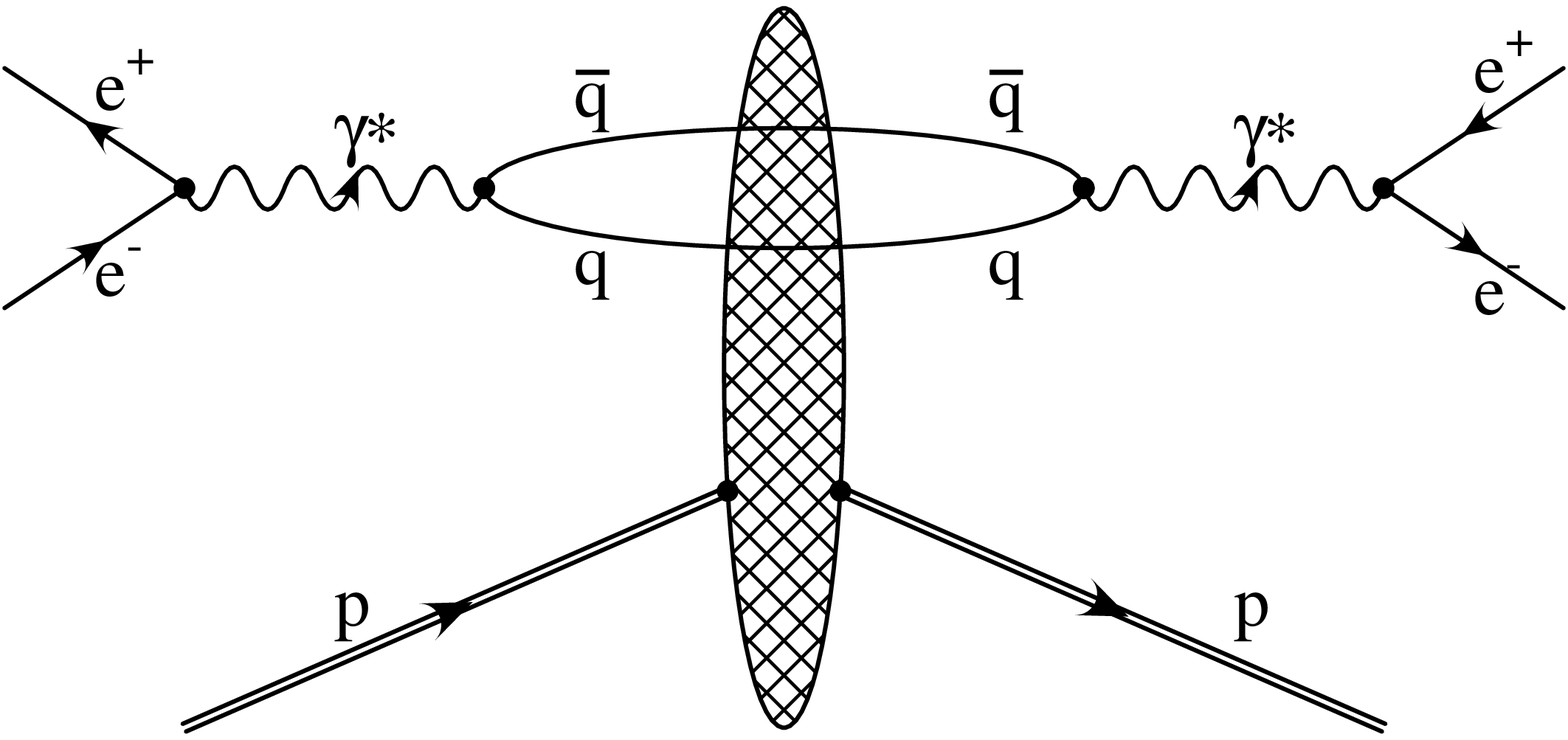,width=10.cm}
 \begin{center} a) \end{center}
\epsfig{file=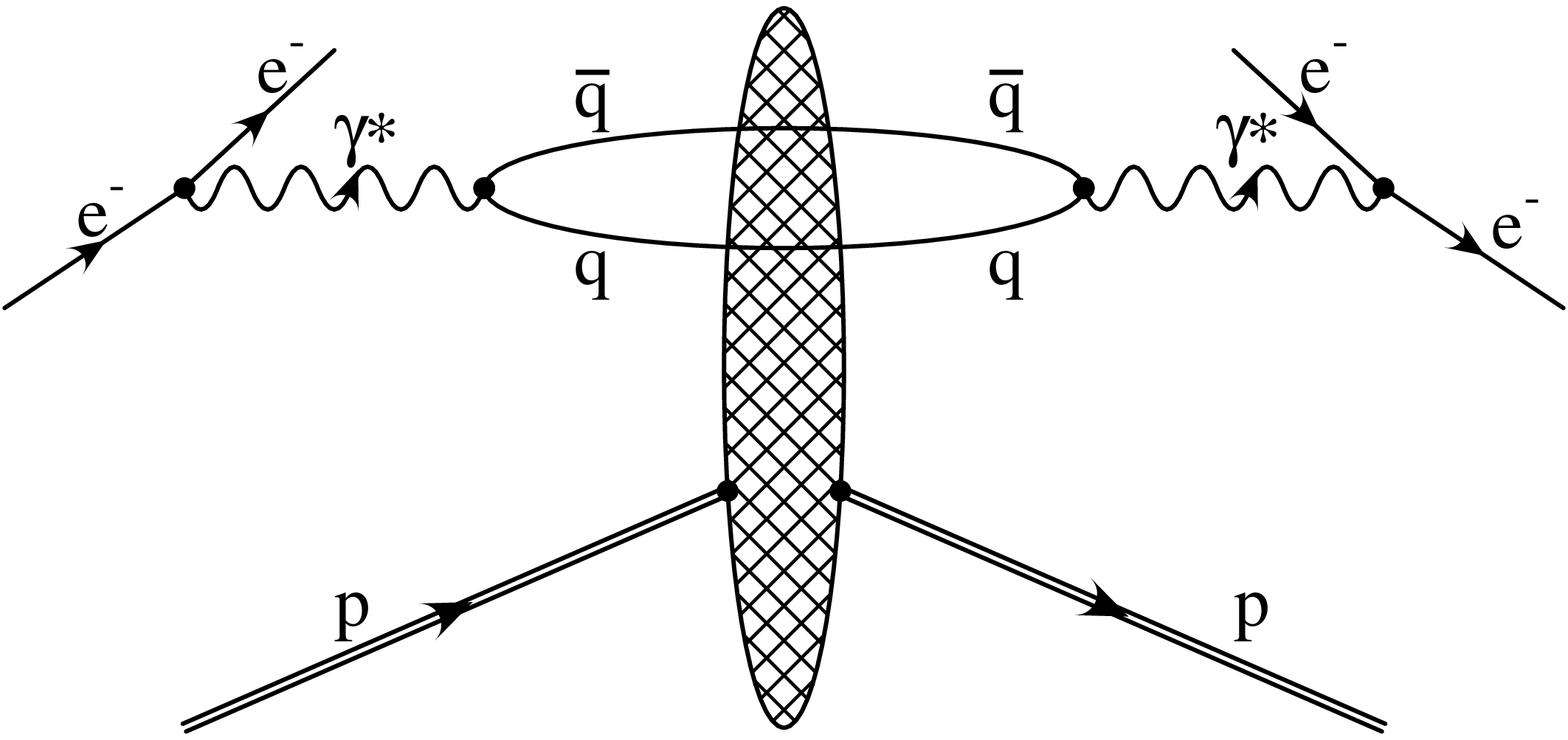,width=10.cm}
 \begin{center} b) \end{center}
\end{center}
\caption{\small The Compton forward amplitude in the proton rest frame, 
a) in the Gedankenexperiment where a timelike photon of mass 
$q^2=M_{q{\bar q}}^2$ interacts with the nucleon, 
b) upon continuation from $q^2=M_{q{\bar q}}^2$ to
$q^2=-Q^2<0$, with $x(\approx Q^2/W^2){\ll}~1$.}
\label{fig2}
\end{figure}

The classical GVD approach is summarized in fig.\ref{fig2}. A virtual timelike photon undergoes a transition to a quark-antiquark, 
$q{\bar q}$, state of mass $M_{q{\bar q}}$.
The $q{\bar q}$ state, upon being 
boosted to high energy in the proton rest frame, is being scattered in the 
forward direction. The appropriate introduction of the propagators for the $q{\bar q}$
  system of mass $M_{q{\bar q}}$ (and of an additional $Q^2$--dependent factor
\cite{ref3, ref15}  from current conservation for longitudinal photon
polarisation) takes us to spacelike four-momentum transfer 
$q^2 \equiv -Q^2 < 0$
with $x \approx Q^2/W^2 \ll 1$ relevant for DIS in the diffraction region.

The configuration of the $q{\bar q}$ vector state depends on the transverse
momentum of the quark, ${\vec k}_{\perp}$, with respect to the three-momentum 
of the photon, and on the (lightcone) variable $z$ ($0 \leq 1 \leq z$) that is
related to the angle, $\vartheta$, of the three momentum of the quark in the $q{\bar q}$-rest-system via
\begin{equation}
{\sin}^{2}\vartheta \equiv 4z(1-z) \ .
\label{eq7}
\end{equation}
The mass of the $q{\bar q}$ system is thus given by\footnote{Here, we work in
  the approximation of massless quarks.}
\begin{equation}
{M}^{2}_{q{\bar q}}=\frac{{k}_{\perp}^{2}}{z(1\!-\!z)} \ ,
\label{eq8}
\end{equation}
such that either the pair of variables $(k^2_{\perp}, z)$ or the pair 
$({M}_{q{\bar q}}, z)$ may be used to characterize the $q{\bar q}$ vector
state. 

Upon carrying out the above-mentioned steps, the transverse and
longitudinal photoabsorption cross sections, 
$\sigma_{\gamma^{\ast}_{T,L}p}(W^2, Q^2)$, become \cite{ref4} 
\begin{eqnarray}
\lefteqn{\!\!\!\!\!\!\!\!\!
{\sigma}_{{\gamma}^{\ast}_{T,L} p}(W^2,Q^2) =
\left[ \frac{1}{2 (2 \pi)^3} \right]^2\!\!\!
\sum_{\lambda, {\lambda}^{\prime}\! =\! \pm 1} 
\int dz \int d{z}^{\prime}\int_{|{\vec k}_{\perp}| 
\ge k_{\perp 0}}\!d^2{k}_{\perp} 
\int_{|{\vec k}_{\perp}^{\prime}| \ge k_{\perp 0}}\!d^2{k}_{\perp}^{\prime}
}
\nonumber\\
&&\!\!\!\!\!\!\!\!\!\!\!\!\!\!\!
\times {\cal M}^{(\lambda,\lambda^{\prime})}_{T,L}( {\vec k}_{\perp}^{\prime},
z^{\prime};Q^2)^{\ast}
 {\cal T}_{(q{\bar q})p \to (q{\bar q})p}({\vec
  k}_{\perp}^{\prime}, z^{\prime}; {\vec k}_{\perp}, z; W^2) 
{\cal M}^{(\lambda,\lambda^{\prime})}_{T,L}({\vec k}_{\perp}, z;Q^2) \ .
\label{eq9}
\end{eqnarray}
The (imaginary part of the) $(q{\bar q})p$ forward-scattering amplitude
(including a factor $1/W^2$ from the use of the optical theorem) has been
denoted by ${\cal T}_{(q{\bar q})p \to (q{\bar q})p}({\vec
  k}_{\perp}^{\prime}, z^{\prime}; {\vec k}_{\perp}, z; W^2)$. The factors 
${\cal M}^{\ast}$ and ${\cal M}$ in (\ref{eq9}) contain the
$\gamma^{\ast}(q{\bar q})$ coupling as well as the propagators and read
\begin{equation}
{{\cal M}}^{({\lambda},{\lambda}^{\prime})}_{T}
({M}_{q{\bar q}},z,Q^2)=
- \frac{{e}_{q}}{Q^2+{M}_{q{\bar q}}^{2}}
\frac{j^{(\lambda,\lambda^{\prime})}_T}{\sqrt{z(1-z)}} \ ,
\label{eq10}
\end{equation}
and 
\begin{equation}
{{\cal M}}^{({\lambda},{\lambda}^{\prime})}_{L}
({M}_{q{\bar q}},z,Q^2)=
- \frac{{e}_{q}}{Q^2+{M}_{q{\bar q}}^{2}}
\sqrt{\frac{Q^2}{M^2_{q{\bar q}}}}
\frac{j^{(\lambda,\lambda^{\prime})}_L}{\sqrt{z(1-z)}} \ . 
\label{eq11}
\end{equation}
We refer to the original paper \cite{ref4} for the explicit expressions for the transverse
and longitudinal currents, $j^{(\lambda,\lambda^{\prime})}_L$ and
$j^{(\lambda,\lambda^{\prime})}_T$. The lower limits of the integrations over
the transverse momenta in (\ref{eq9}) correspond to a finite transverse
extension of the $q{\bar q}$ state in position space (confinement). The
thresholds, $k_{\perp 0}$, allow (\ref{eq9}) to be used as an effective
description at low values of $Q^2$, where the low-lying vector mesons actually
dominate the forward Compton amplitude.

So far, the $(q{\bar q})p$ scattering amplitude has been left unspecified. Quite
generally, transitions diagonal (final mass $M^{\prime}_{q{\bar q}}=M_{q{\bar
    q}}$) and off-diagonal ($M^{\prime}_{q{\bar q}}$ $\neq M_{q{\bar
    q}}$) in the mass of the $q{\bar q}$ state are possible. Assuming $z$ to be 
``frozen'' during the scattering process, diagonal transitions correspond to
zero transverse-momentum transfer to the quark as well as to the antiquark,
${\vec k}_{\perp}^{\prime}={\vec k}_{\perp}$. For off-diagonal transitions to
occur, the same amount of transverse momentum $|{\vec l}_{\perp}|$, opposite in
sign for quark and antiquark, has to be transfered to the system, ${\vec
  k}^{\prime}_{\perp}={\vec k}_{\perp}+{\vec l}_{\perp}$.

\begin{figure*}[htb]
\setlength{\unitlength}{1.cm}
\begin{center}
\epsfig{file=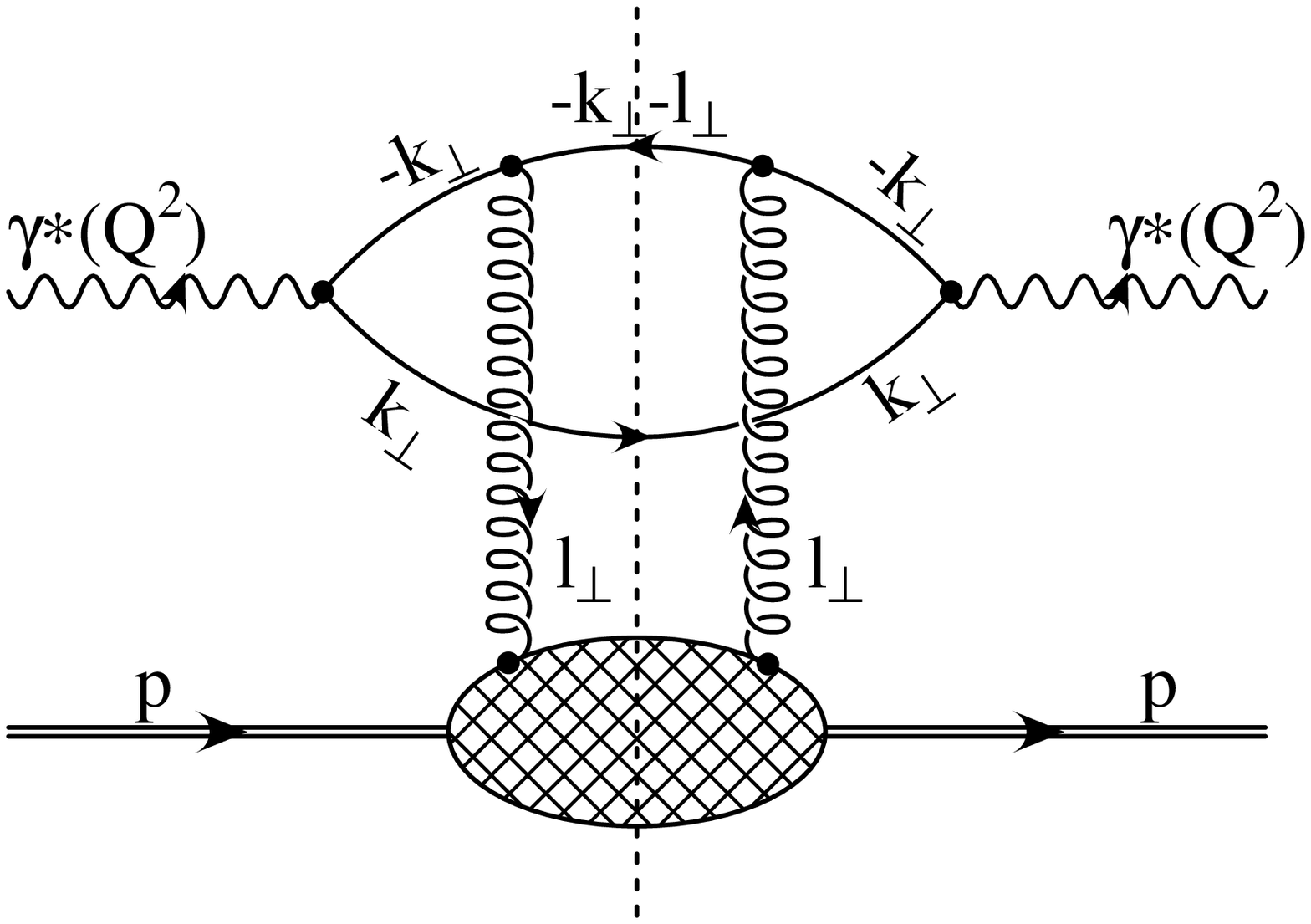,width=7.cm}
 \begin{center} \small a) \end{center}
\vspace{1cm}
\epsfig{file=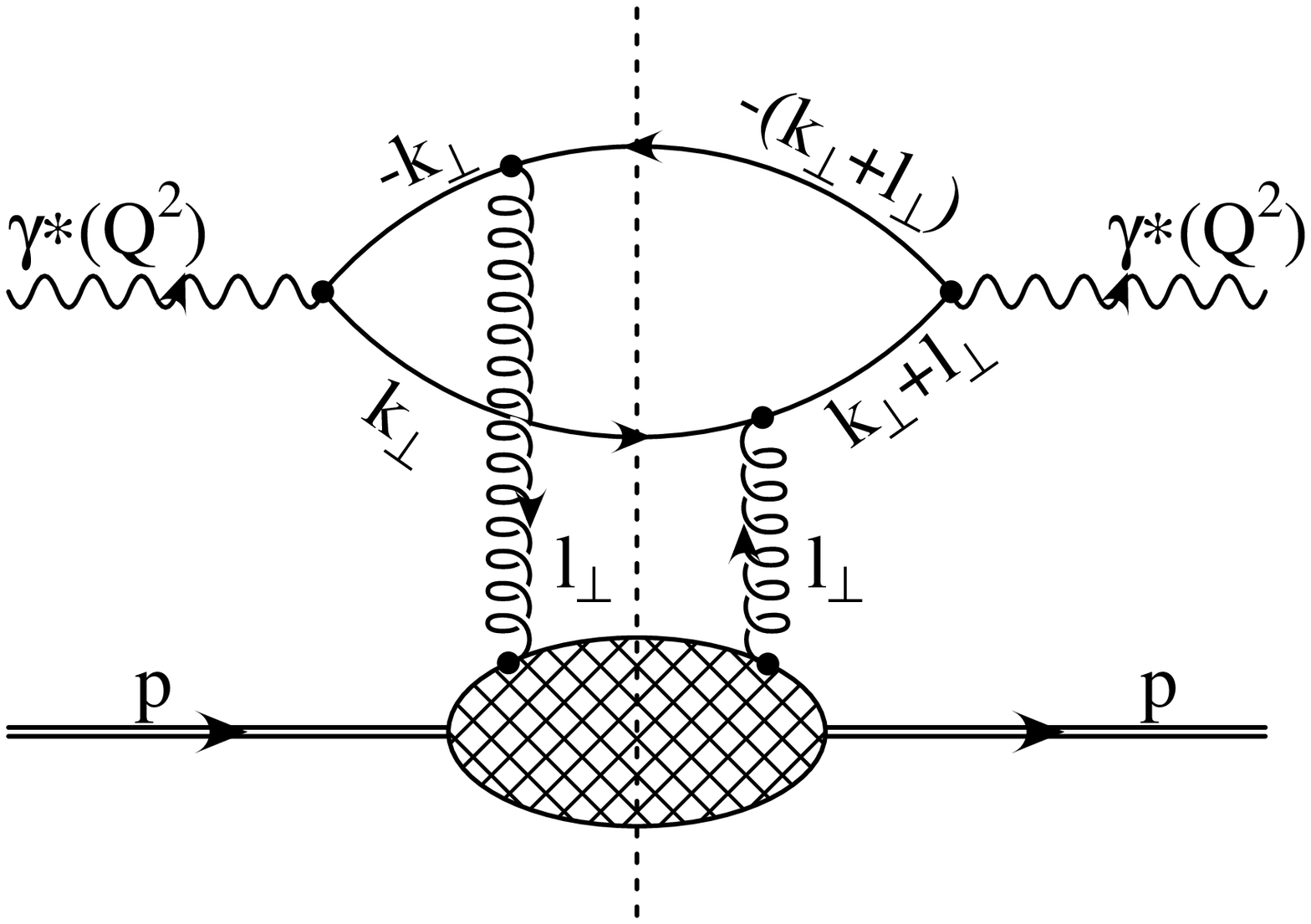,width=7.cm}
 \begin{center} \small b) \end{center}
\caption{\small The two-gluon exchange realisation of the destructive
  interference structure. The diagrams (a) and (b) correspond
  to transitions diagonal and off-diagonal in the masses of the $q{\bar q}$
  pairs, respectively.} 
\label{fig3}
\end{center}
\end{figure*}

It is a general feature of quantum field theory, in particular of QED, or more appropriately for the present case, of pQCD,
that fermion and antifermion couple with opposite sign. This
implies that for every perturbative diagonal transition amplitude there exists
an off-diagonal one of the same strength, but opposite in sign. Compare the two
generic diagrams for the case of two-gluon exchange in fig.\ref{fig3}.
The $(q{\bar q})p$ forward-scattering amplitude is accordingly given by
\begin{eqnarray}
{\cal T}_{(q{\bar q})p \to (q{\bar q})p}({\vec k}_{\perp}^{\prime}, z^{\prime}\!\!&;&\!\!
{\vec k}_{\perp}, z; W^2) = 
2 (2 \pi)^3 \int d^2{l}_{\perp} 
{\tilde \sigma}_{(q{\bar q})p}(l^2_{\perp}, W^2)\times
\nonumber \\
&&
\left \lbrack \delta(
{\vec k}_{\perp}^{\prime}\!-\!{\vec k}_{\perp})\!-\!\delta({\vec
  k}_{\perp}^{\prime}\!-\!{\vec k}_{\perp}\!-\!{\vec l}_{\perp}) \right \rbrack
\delta(z\!-\!z^{\prime}) \ .
\label{eq12}
\end{eqnarray}
Substituting (\ref{eq12}) into (\ref{eq9}), and carrying out the integrations
over $d^2k^{\prime}_{\perp}$ and $dz^{\prime}$, yields
\begin{eqnarray}
&&\!\!\!\!\!\!\!\!{\sigma}_{{\gamma}^{\ast}_{T,L} p}(W^2,Q^2) = 
\frac{1}{16 \pi^3} 
\int dz \int d^2 {l}_{\perp}
{\tilde \sigma}_{(q{\bar q})p}({l}^2_{\perp};W^2)\times \nonumber\\
&&\!\!\!\!\!\!\!\!
{\Bigg \lbrace} \int_{|{\vec k}_{\perp}| \ge k_{\perp 0}} d^2 {k}_{\perp}
\sum_{\lambda, {\lambda}^{\prime} = \pm 1} \left| 
{\cal M}^{(\lambda,{\lambda}^{\prime})}_{T,L} (z,{\vec k}_{\perp}; Q^2) 
\right|^2 -
\nonumber\\
&&\!\!\!\!\!\!\!\!
\int_{|{\vec k}_{\perp}| \ge k_{\perp 0}, |{\vec k}_{\perp}\!+\!{\vec l}_{\perp}| \ge
  k_{\perp 0}}\!\!d^2{k}_{\perp}
\sum_{\lambda, {\lambda}^{\prime} = \pm 1}\!\!
{\cal M}^{(\lambda,{\lambda}^{\prime})}_{T,L} (z,{\vec k}_{\perp}; Q^2) 
{\cal M}^{(\lambda,{\lambda}^{\prime})}_{T,L} 
(z,{\vec k}_{\perp}\!+\!{\vec l}_{\perp}; Q^2)^{\ast}\!
{\Bigg \rbrace}. \nonumber\\ 
\label{eq13}
\end{eqnarray}
The structure of this result, in particular the opposite signs of the diagonal
and the off-diagonal term, coincide\footnote{Compare ref.~\cite{ref4} for 
a more detailed elaboration of this point. 
Similar conclusions, based on somewhat
different reasoning but related arguments, were arrived at in 
refs.~\cite{ref17, ref20, ref19}.} with the structure of destructive
interference embodied in off-diagonal GVD \cite{ref12}. Arguments based on diffraction
dissociation and $q{\bar q}$-bound-state wave functions were orginally employed
to arrive at off-diagonal GVD. Here the original ansatz finds an aposteriori
justification from the general structure of the interaction of a
fermion-antifermion ($q{\bar q}$)
state with a hadron target.

\subsection{The transverse-position-space approach}

The arguments of section \ref{subs1} were based on the classical GVD approach,
characterized by introducing the $Q^2$ dependence via the propagation of 
$q{\bar q}$ vector states. In the present section, I show that identical results
are obtained in a treatment in transverse position space that introduces
QCD in the $(q{\bar q})p$ interaction by invoking the notion of ``colour
transparency'' \cite{ref17, ref16}.

The Fourier transform of the amplitude for the propagating $q{\bar q}$ state, 
${\cal M}^{(\lambda,{\lambda}^{\prime})}_{T,L}$ $(z,{\vec k}_{\perp};Q^2)$, 
from
(\ref{eq10}, \ref{eq11}) yields what is sometimes called the ``photon--$q{\bar
  q}$ wave function'' \cite{ref17},
\begin{equation}
{\psi}_{T,L}^{(\lambda,{\lambda}^{\prime})}(z,{\vec
  r}_{\perp};Q^2){\equiv}\frac{\sqrt{4\pi}}{16{\pi}^{3}}\int_{k_{\perp 0}}
\!d^2{
  k}_{\perp}\exp{({\rm i}{\vec k}_{\perp}\cdot{\vec r}_{\perp})}{\cal
  M}^{(\lambda,{\lambda}^{\prime})}_{T,L}(z,{\vec k}_{\perp};Q^2) \ .
\label{eq14}
\end{equation}
The form of (\ref{eq12}) and (\ref{eq13}) suggests the transverse-position-space ansatz,
\begin{equation}
{\sigma}_{ {\gamma}^{\ast}_{T,L} p }(W^2, Q^2) =
\sum_{\lambda, {\lambda}^{\prime} = \pm 1}
\int dz \int d^2{r}_{\perp} 
\left| \psi_{T,L}^{(\lambda,{\lambda}^{\prime})}(z, {\vec r}_{\perp}; Q^2)
\right|^2 {\sigma}_{(q{\bar q})p}({r}^2_{\perp},W^2)
\ .
\label{eq15}
\end{equation}
According to (\ref{eq15}), the interaction is diagonal in transverse position
space. The notion of colour transparency, no interaction for the $q{\bar q}$
state in the ${r}^2_{\perp} \to 0$ limit of colour neutrality, together
with constancy for ${r}_{\perp}^2$ sufficiently large transverse distance
\begin{equation}
\sigma_{(q{\bar q})p}({r}^2_{\perp}, W^2) \to \left \{ \begin{array}{l}
0, \hspace{2cm} \mbox{for} \quad {r}^2_{\perp} \to 0,\\
\sigma_{(q{\bar q})p}^{(\infty)}(W^2), \quad \mbox{for} \quad {r}^2_{\perp} \to \infty \ ,
\end{array} \right.
\end{equation}
implies the following relation between the ``dipole cross section''
$\sigma_{(q{\bar q})p}({r}^2_{\perp},$ $ W^2)$ and the momentum-space 
function ${\tilde \sigma}_{(q{\bar q})p}({l}^2_{\perp}, W^2)$ of the last
section:
\begin{equation}
{\sigma}_{(q{\bar q})p}({r}_{\perp}^{2},W^2)=\int\!d^2{
  l}_{\perp}{\tilde\sigma}_{(q{\bar q})p}({l}_{\perp}^2,W^2)\left(1-{e}^{-{\rm
  i}{\vec l}_{\perp}\cdot{\vec r}_{\perp}}\right).
\label{eq17}
\end{equation}
Substitution of (\ref{eq17}) and (\ref{eq14}) into (\ref{eq15}) and 
integration over
$d^2r_{\perp}$ and $d^2k^{\prime}_{\perp}$ immediately leads us back to the result (\ref{eq13}) of the last
section. 

The momentum-space GVD approach based on propagators for the $q{\bar q}$ states
of mass $M_{q{\bar q}}$ and a QCD-motivated relative strength of diagonal and
off-diagonal transitions is accordingly entirely equivalent to the transverse
position-space ansatz based on the diagonal representation (\ref{eq15}) and on
the notion of colour transparency.

\begin{figure}[htb]
\setlength{\unitlength}{1.cm}
\begin{center}
\epsfig{file=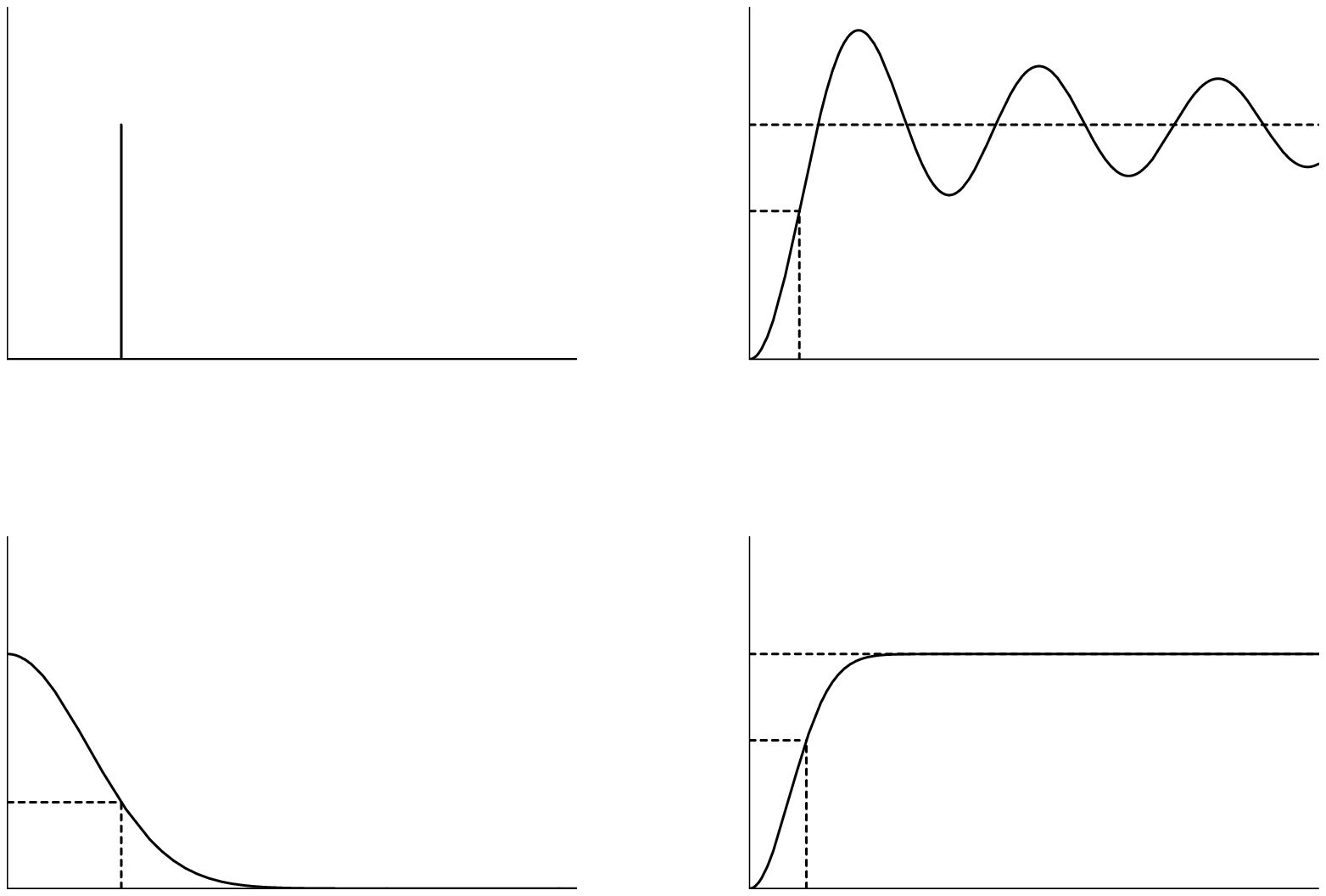, height=16cm, width=12.3cm}
\put(-1,1.2){\makebox(1,1)[c]{${r}_{\perp}$}}
\put(-1,9.2){\makebox(1,1)[c]{${r}_{\perp}$}}
\put(-7.6,1.2){\makebox(1,1)[c]{${l}_{\perp}$}}
\put(-7.6,9.2){\makebox(1,1)[c]{${l}_{\perp}$}}
\put(-12,7.2){\makebox(1,1)[c]{${{\tilde \sigma}_{(q{\bar q})p}({l}_{\perp}^2,W^2)}$}}
\put(-5.3,7.2){\makebox(1,1)[c]{${{\sigma}_{(q{\bar q})p}({r}_{\perp}^2,W^2)}$}}
\put(-5.3,15.2){\makebox(1,1)[c]{${{\sigma}_{(q{\bar q})p}({r}_{\perp}^2,W^2)}$}}
\put(-12,15.2){\makebox(1,1)[c]{${{\tilde \sigma}_{(q{\bar q})p}({l}_{\perp}^2,W^2)}$}}
{\footnotesize
\put(-13,2.9){\makebox(1,1)[c]{$\frac{{\tilde \sigma}(0)}{e}$}}
\put(-6.4,5.2){\makebox(1,1)[c]{$\sigma_{(q{\bar q})p}^{(\infty)}$}}
\put(-6.4,13.1){\makebox(1,1)[c]{$\sigma_{(q{\bar q})p}^{(\infty)}$}}
\put(-11.55,9.2){\makebox(1,1)[c]{$l_{\perp}\!=\!\Lambda$}}
\put(-11.35,1.2){\makebox(1,1)[c]{$l_{\perp}\!=\!1/R_0$}}
\put(-5.15,1.2){\makebox(1,1)[c]{$r_{\perp}\!=\!2 \cdot R_0$}}
\put(-5,9.2){\makebox(1,1)[c]{$r_{\perp}\!=\!1.75/\Lambda$}}
\put(-6.8,3.8){\makebox(1,1)[c]{$0.63 \cdot {\sigma}_{(q{\bar q})p}^{(\infty)}$}}
\put(-6.8,11.8){\makebox(1,1)[c]{$0.63\cdot{\sigma}_{(q{\bar q})p}^{(\infty)}$}}}
\put(-8.7,8.3){\makebox(1,1)[c]{${\small a)}$}}
\put(-8.7,0.5){\makebox(1,1)[c]{${\small b)}$}}
\end{center}
\caption{\small The transverse-position-space dipole cross section
  ${\sigma}_{(q{\bar q})p}({r}^2_{\perp},W^2)$ and its Fourier transform
${\tilde \sigma}_{(q{\bar q})p}({l}^2_{\perp},W^2)$ for two simple choices in
transverse momentum space,  
a) for a ${\delta}$--function and b) for a Gaussian.}
\label{fig4}
\end{figure}

In fig.~\ref{fig4}, as an instructive example, we show the dipole cross section for the simple ansatz of a
$\delta$-function and a Gaussian for ${\tilde \sigma}_{(q{\bar q})p}({l}^2_{\perp}$), 
\begin{eqnarray}
\!\!\!\!\!\!\!\!{\tilde\sigma}_{(q{\bar q})p}({l}_{\perp}^{2})\!&=&\!
\frac{{\sigma}_{(q{\bar q})p}^{(\infty)}}{\pi}
\delta({l}_{\perp}^{2}-{\Lambda}^2) \Rightarrow 
{\sigma}_{(q{\bar q})p}({r}^{2}_{\perp})={\sigma}_{(q{\bar q})p}^{(\infty)}
\left(1\!-\!{J}_{0}({\Lambda}|{\vec r}_{\perp}|)\right) \ ;
\label{eq18}
\\
\!\!\!\!\!\!\!\!{\tilde\sigma}_{(q{\bar q})p}({l}_{\perp}^{2})\!&=&\!
\frac{{\sigma}_{(q{\bar q})p}^{(\infty)}}{\pi}{R}_{0}^{2}
{e}^{-{l}_{\perp}^{2}R_0^2} \Rightarrow  
{\sigma}_{(q{\bar q})p}({r}_{\perp}^{2})={\sigma}_{(q{\bar q})p}^{(\infty)}
\left(1-{e}^{-\frac{{r}_{\perp}^{2}}{4{R}_{0}^{2}}}\right) \ .
\label{eq19}
\end{eqnarray}

Not surprisingly, one finds \cite{ref4} similar results for
$\sigma_{\gamma^{\ast}_{T,L}p}$ in case of the $\delta$-function and the
Gaussian, provided the parameters $\Lambda$ and $R_0$ are related via $\Lambda = 1/R_0$, where $R_0$ is 
of the order of the proton radius, $R_0 \approx 1$ fm $ \approx 0.2 \  
{\rm GeV}^{-1}$. A Gaussian ansatz was employed in a recent analysis
\cite{ref18} of the experimental data, while a polynomial representation was
used in ref.~\cite{ref19}.

\subsection{A remark on the two-gluon exchange}

In the above treatment, the two-gluon exchange only entered in terms of a
generic structure of the interaction of the fermion-antifermion ($q{\bar q}$) 
state with the
nucleon. As a cross-check, one may alternatively evaluate the two-gluon 
exchange diagrams in the low-$x$ limit of DIS as a more concrete QCD model. In order to treat the low-$x$ limit, the
introduction of Sudakov variables proves useful \cite{ref17, ref17b}. As long as no detailed model
assumptions are introduced for the lower vertex in fig.~\ref{fig3}, the final
result coincides with the one obtained above, even though ${\tilde
  \sigma}_{(q{\bar q})p}({l}^2_{\perp}, W^2)$ now contains the quark-gluon coupling,
colour factors and the gluon propagator. The treatment is of interest,
nevertheless, as in this covariant approach, one does not rely on the propagator
rule for the $q{\bar q}$ system or the colour transparency assumption. In the
sense of leading to the same result as the $q{\bar q}$-state propagator rule,
the structure of off-diagonal GVD is realized by the pQCD model of two-gluon
exchange. The validity of the expression (\ref{eq13}) for
$\sigma_{\gamma^{\ast}_{T,L}p}$ is more general, however, not strictly
bound to the two-gluon exchange ansatz.

\section{The $Q^2$ dependence explicitly}

To obtain the $Q^2$ dependence contained in the expression for
$\sigma_{\gamma^{\ast}_Tp}(W^2,$ $ Q^2)$ in (\ref{eq13}) explicitly, ${\tilde
  \sigma}_{(q{\bar q})p}$ must be specified, and a fivefold integration
is to be carried out. I will restrict myself to the main steps and refer to the
original paper \cite{ref4} for a detailed presentation of the procedure.

Carrying out the angular integration over the direction of the transverse
momentum of the incoming quark, ${\vec k}_{\perp}$, in (\ref{eq13}), and introducing the
masses of the $q{\bar q}$ system appearing in the propagators, one remains with a
fourfold integration over $dz dl_{\perp}^2 dM^2_{q{\bar q}} dM^{\prime 2}_{q{\bar
    q}}$. Concerning ${\tilde \sigma}_{(q{\bar q})p}(l^2_{\perp}, W^2)$, for simplicity,
in the recent paper, we refrained from introducing an appropriate $W$
dependence, surely necessary for a precise description of the experimental
data. We rather carried out simple model calculations based on the
$\delta$--function and the Gaussian ansatz in (\ref{eq18}) and (\ref{eq19}). The
essential conclusions on the origin of the $Q^2$ dependence from $q{\bar q}$-
vector-state propagation, arrived at with this procedure,
will remain valid in a more complete treatment, also taking into account the
energy dependence.

Even upon specializing (\ref{eq13}) to the $\delta$--function ansatz (\ref{eq18}), a complete
analytical treatment of the remaining threefold integral cannot be carried out
for arbitrary values of $Q^2$. Nevertheless, simple and practically exact
expressions for the $Q^2$ dependence of $\sigma_{\gamma^{\ast}_{T,L}p}(W^2,
Q^2)$ can be derived by applying the mean-value theorem to the integrations over
the configuration variable $z$ and the mass $M^{\prime 2}_{q{\bar q}}$. 

In the expressions
for the transverse cross section (\ref{eq13}), we replace $z$, or rather
$z(1-z)$, by the mean value
\begin{equation}
\kappa_T=\kappa_T(Q^2)={\bar z}_T(Q^2)(1-{\bar z}_T(Q^2)) \ .
\label{eq20}
\end{equation}
We have indicated a potential $Q^2$ dependence of ${\bar z}_T(Q^2)$. Likewise,
we replace $M^{\prime 2}$ by an average, conveniently parametrized by the
parameter $\delta_T$ according to
\begin{equation}
{\overline M^{\prime 2}}= M^2+
\frac{{l}_{\perp}^{2}}{{\bar z}(1\!-\!{\bar z})}\frac{1}{(1\!+\!2 {\delta}_T)} \ .
\label{eq21}
\end{equation}
After these replacements, the integration over $dM^2$ can be carried out to
yield \cite{ref4}
\begin{eqnarray}
&&\!\!\!\!\!{\sigma}_{ {\gamma}^{\ast}_T p }(W^2, Q^2; {\kappa}_T(Q^2), \delta_T) 
= \frac{\alpha}{2 \pi} \left( \frac{e_q}{e_0}
\right)^2 {\sigma}^{(\infty)}_{(q{\bar q})p}(1-2{\kappa}_T(Q^2))
\times
\nonumber\\
&&\!\!\!\!\!\!\!\!
{\Bigg \lbrack}\!\!\! 
\left (\!
(1\!+\!2 \delta_T) \frac{Q^2}{{\Lambda}^{2}}\kappa_T(Q^2)+(1\!+\!\delta_T)\!\right)
\ln \left (\!\!1\!+\!\frac{{\Lambda}^{2}}{\kappa_T(Q^2)(1\!+\!2 \delta_T)
(Q^2\!+\!M_0^2({\kappa}_T(Q^2)))}\!\!\right )
\nonumber\\
&&
- \frac{Q^2}{ (Q^2+M_0^2({\kappa}_T(Q^2))) } 
{\Bigg \rbrack} \ .
\label{eq22}
\end{eqnarray} 
The mass $M_0(\kappa_T(Q^2)) \equiv k_{\perp 0}^2/\kappa_T(Q^2)$ corresponds to
the threshold $k_{\perp 0}^2$ introduced in (\ref{eq9}). In addition to the
basic physical parameters, $\sigma_{(q{\bar q})p}^{(\infty)}$, normalizing the
cross section, and $\Lambda^2$ determining the value of the three-momentum,
$l_{\perp}^2$, transferred to the $q{\bar q}$ system, the expression for
$\sigma_{\gamma^{\ast}_Tp}$ in (\ref{eq22}) now contains the theoretical
quantities $\kappa_T(Q^2)$ and $\delta_T$. Their numerical values are obtained by comparing the
mean-value result 
(\ref{eq22}) with the exact result obtained by numerical integration of the 
basic expression (\ref{eq13}). While $\delta_T$ turns out to be independent 
of $Q^2$,
the configuration, $\kappa_T(Q^2)$, is indeed found to depend on $Q^2$ logarithmically,
\begin{equation}
\kappa_T(Q^2)=\frac{3}
{6+4 \ln \left (c_1 \frac{Q^2}{\Lambda^2}+\exp(c_2) \right )} \ .
\label{eq23}
\end{equation}
The constant $c_2$ is uniquely determined by the mean-value evaluation of the
photoproduction cross section, that yields $\kappa_T(0)$, while $c_1$ is deduced
from the asymptotic $Q^2$ behaviour.
Asymptotically, from (\ref{eq22}) with $\kappa_T(Q^2)$ from (\ref{eq23}), one obtains
\begin{equation}
{\sigma}_{ {\gamma}^{\ast}_T p }(W^2, Q^2\!\to\!\infty) 
\!=\!\frac{{\alpha}}{3 \pi}\left( \frac{e_q}{e_0} \right)^2
\!{\sigma}^{(\infty)}_{(q{\bar q})p}\!\left \lbrack\!\! 
\frac{{\Lambda}^2}{Q^2} \ln \!\left (\frac{Q^2}{{\Lambda}^2} \!\right )\!+\! 
(\ln c_1)\frac{\Lambda^2}{Q^2}\!+\!{\cal O} \left ( \frac{\ln Q^2}{Q^4} \right
 )\!\!
\right \rbrack.
\label{eq24}
\end{equation}
For the ensuing discussion of the $(1/Q^2) \log Q^2$ dependence in (\ref{eq24}),
it is useful to also present the asymptotic result obtained from (\ref{eq22}),
if we ignore the $Q^2$ dependence of $\kappa_T(Q^2)$ by putting $\kappa_T(Q^2)
\equiv \kappa_T(Q^2=0)$. In this case, one finds,
\begin{equation}
{\sigma}_{ {\gamma}^{\ast}_T p }(W^2, Q^2\!\to\!\infty; {\kappa}_T(0)) 
= \frac{\alpha}{3 \pi} \left( \frac{e_q}{e_0} \right)^2 
{\sigma}^{(\infty)}_{(q{\bar q})p} \frac{3}{4}
\frac{(1\!-\!2{\kappa}_T(0))}{ {\kappa}_T(0)}
\left \lbrack \frac{{\Lambda}^{2}}{Q^2}\!+\!{\cal O}\left
 (\frac{1}{Q^4}\right )\!\!
\right \rbrack ,
\label{eq25}
\end{equation}
i.e. by ignoring the $Q^2$ dependence of the configuration, one loses the $\log
Q^2$ factor present\footnote{The asymptotic $Q^2$ dependence (\ref{eq24}) was
  also verified analytically \cite{ref4} directly from (\ref{eq13}) without
  application of the mean-value theorem.} in (\ref{eq24}). We conclude that the
logarithmic $Q^2$ dependence of the $q{\bar q}$ configuration, implicitly
contained in (\ref{eq13}), is essential for the $\log Q^2$ dependence in
(\ref{eq24}) that corresponds to a logarithmic violation of scaling for the
structure function $F_2$. The example of putting $\kappa_T(Q^2) \equiv
\kappa_T(0)$ demonstrates that the $1/Q^2$ (scaling) dependence is unaffected 
by ignoring the $Q^2$ dependence of the configuration. In fact, as repeatedly
stressed, it is the destructive interference between diagonal and off-diagonal
transitions, characteristic for off-diagonal GVD, that leads to the $1/Q^2$
scaling behaviour. This conclusion is substantiated also by the results for 
the longitudinal cross section to be given below. In the longitudinal case, the
configuration variable turn out to be independent of $Q^2$ in conjunction with a
$1/Q^2$ scaling of the cross section.

\begin{figure}[htb]
\setlength{\unitlength}{1.cm}
\begin{center}
\epsfig{file=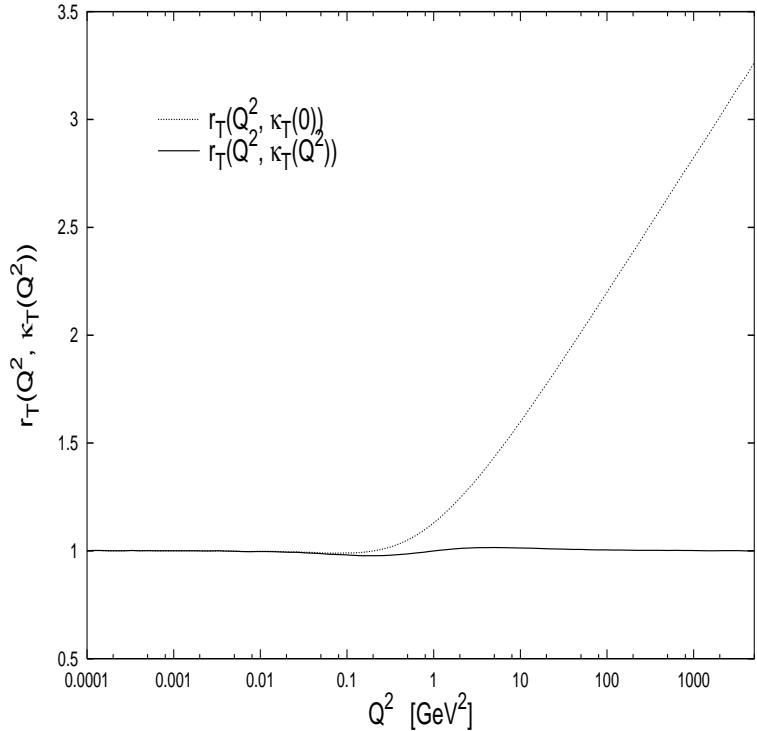, height=10cm, width=10cm}
\end{center}
\caption{\small The lines show the ratio 
$r_T(Q^2, \kappa_T(Q^2))={\sigma}_{{\gamma}^{\ast}_Tp}(W^2, Q^2)/
{\sigma}_{{\gamma}^{\ast}_Tp}(W^2, Q^2; {\kappa}_T(Q^2), \delta_T)$ from 
(\ref{eq26}). The numerator is obtained by numerical integration of
(\ref{eq13}), the denominator by evaluating the mean-value expression
(\ref{eq22}). The solid line is based on $\kappa_T(Q^2)$ from (\ref{eq23}). The dotted line shows the result of ignoring the $Q^2$ dependence of
$\kappa_T(Q^2)$ by putting $\kappa_T(Q^2) \equiv \kappa_T(0)$ in
(\ref{eq22}).}
\label{fig5}
\end{figure}

In order to explicitly demonstrate the validity of the mean-value evaluation (\ref{eq22}), in
fig.~\ref{fig5}, we present the ratio $r_T(Q^2, \kappa_T(Q^2))$,
\begin{equation}
r_{T}(Q^2, \kappa_T(Q^2)) \equiv
\frac{{\sigma}_{{\gamma}_{T}^{\ast}p(W^2, Q^2)}}
{{\sigma}_{{\gamma}_{T}^{\ast}p(W^2, Q^2; {\kappa}_T(Q^2), \delta_T)}} \ .
\label{eq26}
\end{equation}
The numerator in (\ref{eq26}), the exact numerical result for $\sigma_{\gamma^{\ast}_{T,L}p}$, is obtained by numerical integration of
(\ref{eq13}). The denominator is based on the mean-value result
(\ref{eq22}). For the numerical evaluation, we chose $\Lambda^2/k_{\perp 0}^2=1$
and $\Lambda^2=0.05$. For this choice of $\Lambda^2$ we found for the
theoretical parameters $\delta_T$ in (\ref{eq21}) and $c_1$, $c_2$ in
(\ref{eq23}), 
\begin{equation}
\delta_T=0.52, \quad c_1=1.50, \quad c_2=3.65 \ .
\label{eq27a}
\end{equation}
The value of $c_2=3.65$ corresponds to $\kappa_T(0)=0.1455$. As seen in
fig.~\ref{fig5}, with $\kappa_T(Q^2)$ from (\ref{eq23}), the mean-value
evaluation practically coincides with the exact result: $r_T(Q^2,
\kappa_T(Q^2))=1$ is well fulfilled for all $Q^2 \ge 0$. 
\begin{table}
\begin{center}
\begin{tabular}{|c|c|c|c|}\hline
$Q^2 \ [{\rm GeV}^2]$ & $\kappa_T(Q^2)$ & $\sin \vartheta$ & $\vartheta$ \\ \hline
$0.01$      & $0.1453$ & $0.76$ & $49.46^0$ \\ \hline
$1$         & $0.1309$ & $0.72$ & $41.25^0$ \\ \hline
$100$       & $0.0788$ & $0.56$ & $32.09^0$ \\ \hline
\end{tabular} 
\end{center}
\caption{The parameter $\kappa_{T}(Q^2)$ from (\ref{eq23}) and the related 
angular dependences as a function of $Q^2$. We used ${\Lambda^2}/k^2_{\perp 0}=1$, or $k^2_{\perp 0}=0.05 \ {\rm GeV}^2$.}
\label{table}
\end{table}    

Dropping the $Q^2$ dependence by replacing $\kappa_T(Q^2)$ by
$\kappa_T(0)$ in (\ref{eq22}) and (\ref{eq26}), as expected from the above discussion,  as a consequence of
the missing $\log Q^2$ term in (\ref{eq25}), implies a linear
asymptotic rise in $\log Q^2$ for $r_T(Q^2, \kappa_T(0))$. Compare fig.~\ref{fig5}.
In table \ref{table}, we show a few numerical values for $\kappa_T(Q^2)$. 
Its slow logarithmic dependence is responsible for the $\log Q^2$
factor in $\sigma_{\gamma^{\ast}_Tp}$, corresponding to a logarithmic violation
of scaling for $F_2$.

An analogous procedure may be applied to the longitudinal cross section,
$\sigma_{\gamma^{\ast}_Lp}$. In distinction from the transverse case, it turns
out that both, $\kappa_L \equiv {\bar z}_L(1\!-\!{\bar z}_L)$ and $\delta_L$ are
independent of $Q^2$. The longitudinal cross section in mean-value evaluation becomes
\begin{eqnarray}
\lefteqn{
{\sigma}_{ {\gamma}^{\ast}_L p }(W^2, Q^2; {\kappa}_L, \delta_L) 
=\frac{2 \alpha}{\pi} 
\left( \frac{e_q}{e_0}\right)^2 {\sigma}^{(\infty)}_{(q{\bar q})p} Q^2 
\kappa_L
{\Bigg \lbrack} \frac{1}{(Q^2+M_0^2({\kappa}_L))}  
} \nonumber\\
&&
- \frac{ (1\!+\!2 \delta_L)\kappa_L}{ {\Lambda}^{\prime 2} }
\ln \left( 1+\frac{ {\Lambda}^{\prime 2} }
{ \kappa_L(1\!+\!2 \delta_L)(Q^2+M_0^2({\kappa}_L)) } \right )
{\Bigg \rbrack} \ .
\label{eq27}
\end{eqnarray} 
For $Q^2 \to \infty$ we have a $1/Q^2$ dependence, corresponding to a scaling
contribution to $F_2$,
\begin{eqnarray}
{\sigma}_{ {\gamma}^{\ast}_L p }(W^2, Q^2 \to \infty; \delta_L) 
= \frac{2 \alpha}{{\pi}} 
\left( \frac{e_q}{e_0} \right)^2 {\sigma}^{(\infty)}_{(q{\bar q})p} 
{\Bigg \lbrack} 
\frac{{\Lambda}^2}{2 (1\!+\!2 \delta_L)Q^2}
\!+\!{\cal O}\left(\frac{1}{Q^4} \right)
{\Bigg \rbrack},
\label{eq28}
\end{eqnarray}
while for $Q^2 \to 0$ the longitudinal cross section rises linearly with
$Q^2$. The numerical values for $\kappa_L$ and $\delta_L$ are given by
\begin{equation}
\kappa_L=0.1714, \quad \delta_L=-0.1767 \ .
\label{eq29}
\end{equation}
As in the transverse case, the mean-value evaluation (\ref{eq27}) provides an
excellent approximation of the exact result obtained by numerical integration 
of (\ref{eq13}). Compare the original paper \cite{ref4}.

\begin{figure}[htb]
\setlength{\unitlength}{1.cm}
\begin{center}
\epsfig{file=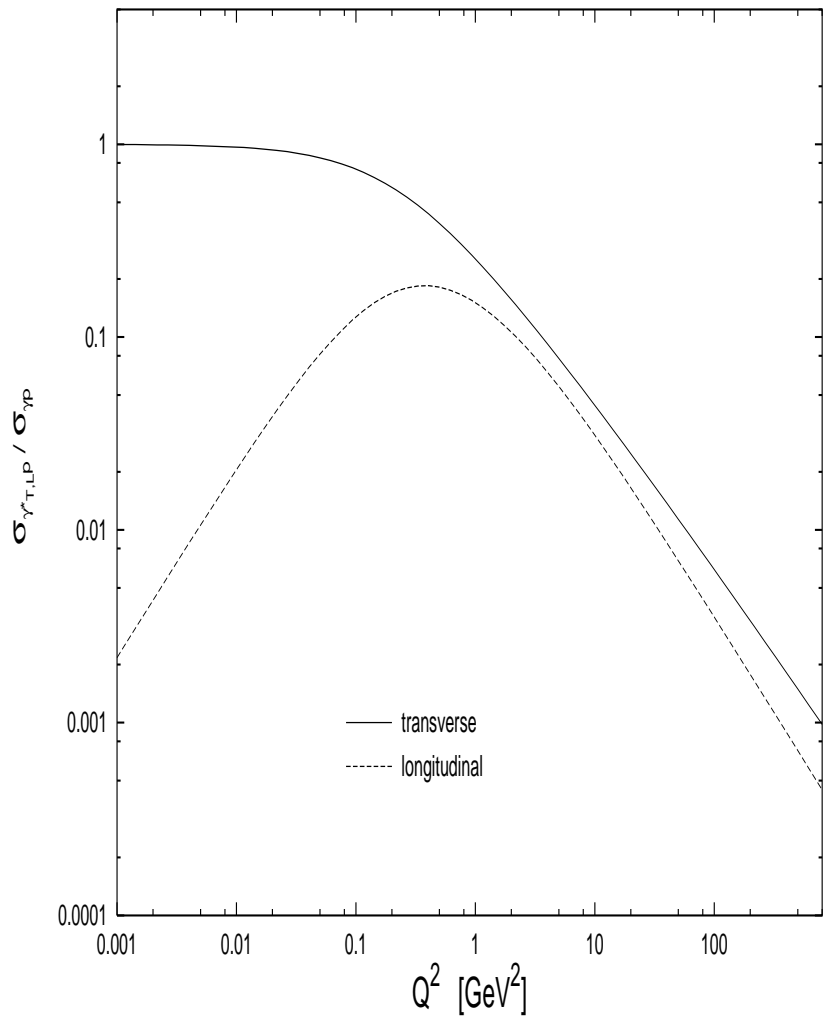, height=14cm, width=11cm}
\end{center}
\caption{\small Numerical results for ${\sigma}_{ {\gamma}^{\ast}_Tp }$
(solid line), and ${\sigma}_{{\gamma}^{\ast}_Lp }$ (dotted
line) from (\ref{eq13}), normalized by the
photoproduction cross section ${\sigma}_{{\gamma}p}$. The results shown are
obtained by numerical integration of (\ref{eq13}). The
mean-value results (\ref{eq22}) (with $\kappa_T(Q^2)$ from (\ref{eq23})) and
(\ref{eq27}) coincide with the ones shown, apart from a minor
deviation in the longitudinal cross section around $Q^2 \approx 1 \ {\rm
  GeV}^2$.} 
\label{fig7}
\end{figure}

In fig.~\ref{fig7}, we show the transverse and longitudinal cross sections
normalized by (transverse) photoproduction. 
For a detailed comparison with
experiment, the theory has to be extended by incorporating the $W$
dependence. Compare refs. \cite{ref19, ref18} for interesting work in this direction. Nevertheless, it is worth noting that the drop by almost $2$ orders
of magnitude from $Q^2 = 0 \ GeV^2$ to $Q^2 \approx 100 \ GeV^2$ seen in the experimental data in fig.~\ref{fig8} is also present 
in the present theoretical results.

\begin{figure}
\setlength{\unitlength}{1.cm}
\begin{center}
\epsfig{file=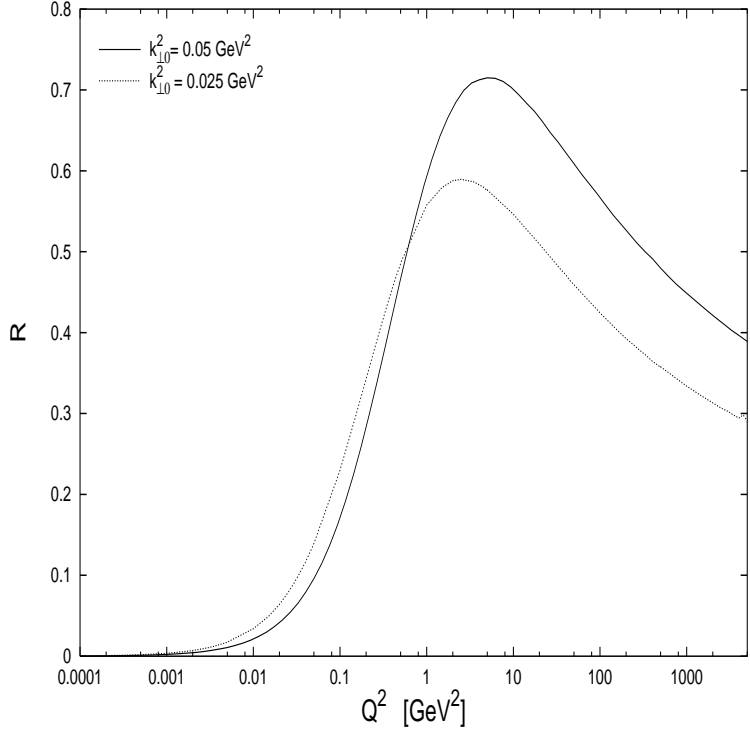, height=10cm, width=10cm}
\end{center}
\caption{\small The longitudinal-to-transverse ratio
  $R$. The solid curve corresponds to 
${\Lambda^2}=k_{\perp 0}^2=0.05 \ {\rm GeV}^2$ as used in
Fig.~\ref{fig7}. The dotted curve is obtained for $\Lambda^2=0.05 \ {\rm
  GeV}^2$ and $k_{\perp 0}^2=0.025 \ {\rm GeV}^2$, as indicated.}
\label{fig9}
\end{figure}

Finally, in fig.~\ref{fig9} we show the longitudinal to transverse ratio $R$. 
It rises linearly for $Q^2 \to 0$ and drops as $1/\log Q^2$ for $Q^2 \to
\infty$.

\section{Off-diagonal GVD in vector-meson production}
 
Reformulating and extending the off-diagonal GVD ansatz \cite{ref23} 
for elastic vector
meson production, recent work \cite{ref21} by Schuler, Surrow and myself
yields a satisfactory representation of the transverse cross section
and the longitudinal-to-transverse ratio, $R$, for elastic $\rho^0, \phi$
and $J/{\rm Psi}$-production \cite{ref21, ref24}. The theoretical 
prediction for $\sigma_{T, \gamma^*p \to Vp}$ is based on
\begin{figure}[htb]
\begin{center}
\setlength{\unitlength}{1cm}
\begin{minipage}[h]{6cm}
\begin{picture}(6,13)
\put(-1.5,0){\includegraphics{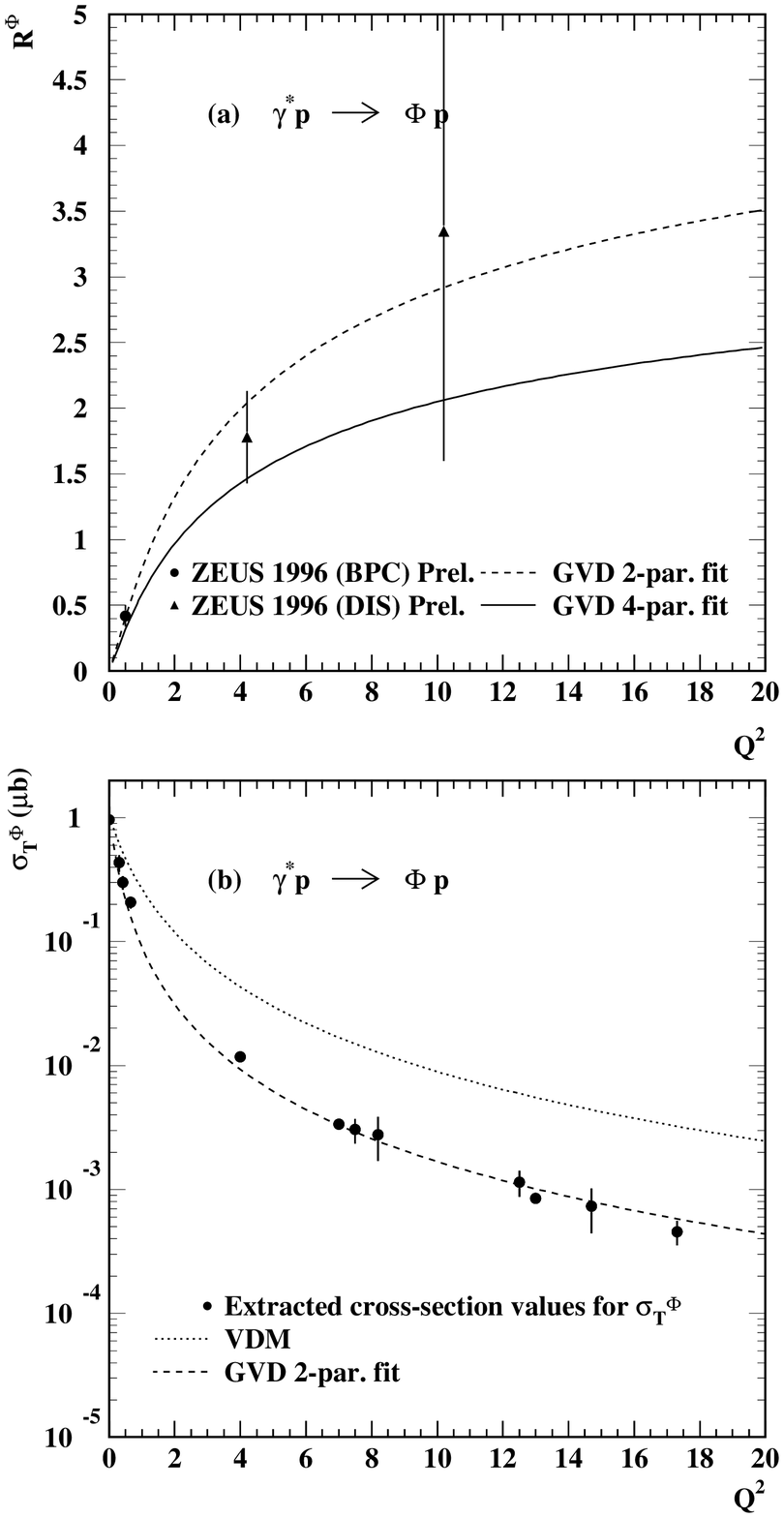}}
\end{picture}\par
\caption{GVD in $\gamma^*p \to \phi~ p$ \cite{ref21}.}
\label{fig11}
\end{minipage}\hfill
\begin{minipage}[t]{6cm} 
\begin{picture}(6,6)
\put(-0.5,-1){\includegraphics{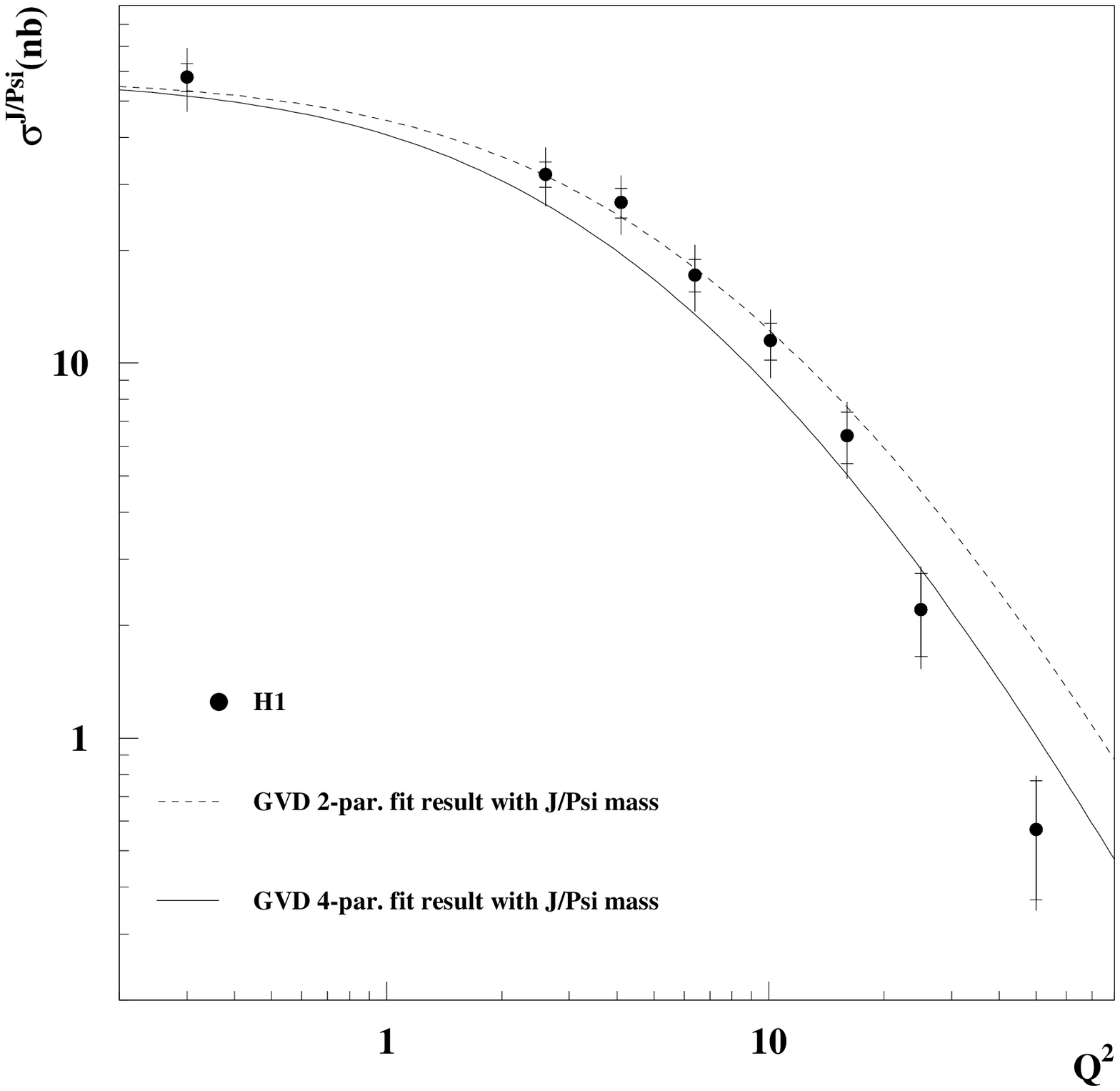}}
\end{picture} 
\caption{GVD in $\gamma^*p \to J/{\rm Psi}~ p$ \cite{ref24}.}
\label{fig12}
\end{minipage}
\end{center}
\end{figure}
%
%
\begin{equation}
\sigma_{\gamma^*_T p \to V p} = \frac{m^4_{V,T}}{(Q^2 + m^2_{V,T})^2} \sigma_{\gamma p \to V p} (W^2) \ .
\end{equation}
I refer to ref.~\cite{ref21} for the prediction for $R$. The inclusion of
off-diagonal transitions with destructive interference yields $m^2_{V,T}
< m^2_V$, where $m_V$ stands for the mass of the vector meson being
produced. As an example, in fig.~\ref{fig11}, I show $\phi$ production. The curves
are based on $m^2_{\phi,T} = 0.40 m^2_\phi$ and 
$\sigma_{\gamma p \to \phi p} = 1.0 \mu b$. The theoretical curves for
$J/{\rm Psi}$ production in fig.~\ref{fig12} were obtained by the replacement $m^2_\phi
\to m^2_{J/{\rm Psi}}$ and $\sigma_{\gamma p \to \phi p} \to \sigma_{\gamma p
\to J/{\rm Psi}~ p}$.
A more direct connection between the result of ref.~\cite{ref4} and the
off-diagonal GVD treatment in ref.~\cite{ref21} needs to be established.

\section{Conclusions}

I will be brief in my conclusions. Taking into account the $q{\bar q}$
configuration in the $\gamma^{\ast} \to q{\bar q}$ transition and the opposite 
signs of the fermion and the antifermion interaction with the nucleon target in
the formulation of GVD, one arrives at a representation of the photoabsorption
cross section containing destructive interference between diagonal and
off-diagonal transitions characteristic for off-diagonal GVD. The destructive
interference is responsible for scaling, the implicit dependence on the 
$q{\bar q}$ configuration is responsible for the logarithmic violation of
scaling. The longitudinal-to-transverse ratio decreases logarithmically with $Q^2$ for $Q^2 \to \infty$. The
classical momentum-space formulation has been shown to lead to results
identical to the position-space treatment based on the notion of colour
transparency. Needless to be stressed again, off-diagonal GVD is compatible with
QCD, it is in fact contained in QCD. The two-gluon exchange provides the
simplest QCD model showing the features of the more general GVD-momentum-space
or the transverse-position-space formulation. Off-diagonal GVD also yields 
the experimentally observed $Q^2$
dependence for (elastic) vector-meson forward production. A careful treatment of
the energy dependence, and further work on the diffractively produced final states
will be indispensable in order to approach a detailed understanding of
deep-inelastic scattering in the low-x diffraction region.

\section*{Acknowledgement}
It is a pleasure to thank my Polish colleagues and friends for the organization
of a very successful meeting and a pleasant stay in Ustron.

\setlength{\unitlength}{1cm}


\end{document}